\documentclass[mathlef]{an}
\usepackage{graphicx}
\usepackage{times}
\begin{document}

\Pagespan{789}{}
\Yearpublication{2008}%
\Yearsubmission{2008}%
\Month{11}%
\Volume{999}%
\Issue{88}%

\title{Structure, spectra and variability of some GPS radio
sources}

\author{X. Liu\inst{1}\fnmsep\thanks{Corresponding author:
  \email{liux@uao.ac.cn}\newline}
\and  H. -G. Song\inst{1} \and  L. Cui\inst{1,2} }
\titlerunning{Structure, spectra and variability of GPS
sources}
\authorrunning{X. Liu et al.}
\institute{ National Astronomical Observatories/Urumqi
observatory, CAS, 40-5 South Beijing Rd, Urumqi 830011, PR China
\and Graduate University of the Chinese Academy of Sciences,
Beijing 100049, PR China}

\received{2008} \accepted{ 2008} 


\abstract{ We report the results of multifrequency-VLBI
observations of GHz-Peaked-Spectrum (GPS) radio sources. The VLBI
structure and component spectra of some GPS sources are presented.
Our VLBI results show that about 80\% of the GPS galaxies exhibit
a compact double or CSO-like structure, while the GPS quasars tend
to show a core-jet. The component spectra of the GPS galaxies are
often steep/convex, and the core has a flat spectrum but it is
usually hidden or weak. In addition, we studied the variability of
GPS sources by comparing new flux density measures, acquired with
the Urumqi 25m telescope at 4.85 GHz, with previous 87GB data. The
results show that 44\% of the GPS quasars varied higher than 10\%
in passed 20 years, while the fraction is only 12\% for the GPS
galaxies meaning that the GPS quasars are much more variable than
GPS galaxies. In total, 25\% of GPS sources show $>10\%$
variability at 4.85 GHz in our sample.}

\maketitle

\section{Introduction}

GHz-Peaked-Spectrum (GPS) radio sources are powerful
($P_{\rm1.4~GHz}\geq\rm10^{25}\,W\,Hz^{-1}$) and compact
($\leq1$~kpc), characterized by a convex radio spectrum that peaks
between 0.3-10~GHz (observer's frame), and they represent a
significant fraction ($\sim$ 10\%) of the bright radio source
population (O'Dea 1998). Only a few percent of them have weak
extended emission (Stanghellini et al. 2005). A couple of GPS
sources are also identified as Compact Symmetric Objects (CSOs,
e.g Owsianik \& Conway 1998). It is believed that GPS-small sizes
are most likely due to their youth ($<{10^4}$ years) than to a
dense confining medium (Murgia et al. 1999; O'Dea et al. 2005). In
this scenario, these sources will evolve into large radio sources
($>15$~kpc), some of them would evolve into FRII radio sources
(Fanti et al. 1995, Snellen et al. 2000). However, the GPS
phenomenon is not completely understood, e.g. the distribution of
sizes/ages, structure/spectra, and variability of GPS sources. We
have carried out EVN (European VLBI Network) observations of 19
GPS sources, 15 of them are from the Parkes half-Jansky sample
(Snellen et al. 2002) with declination $>-5^\circ$ and not
observed with VLBI before (except 2121$-$014 and 2322$-$040 we
observed), four others are from our previous observation list
which observed with the EVN at 2.3/8.4 GHz and/or 5 GHz (Xiang et
al. 2005, 2006), the 1.6 GHz observation will further provide
information on the source structure and spectra. Furthermore, we
have carried out flux density observations with the Urumqi 25m
telescope in order to study GPS flux density variability,
searching for a different behavior between galaxies and quasars.

  \begin{table*}

         \caption[]{The GPS sources. Columns (1),(2) source names; (3) optical identification (gl: galaxy, Q: quasar, EF: empty field); (4) optical
         magnitude; (5)
         redshift (de Vries et al. 2007, those with * are photometric estimated by Tinti et al. 2005); (6) linear scale factor pc/mas
         [$H_{0}=71 km s^{-1} Mpc^{-1}$ and $q_{0}=0.5$ have been
         assumed]; (7) maximum angular size from the observation; (8) maximum linear size; (9)
         VLBI structure (cd: compact double, cj: core-jet, n: no detection); (10) low frequency spectral index; (11) higher frequency spectral index; (12) turnover frequency; (13)
peak flux density; (14) references for the spectral information, 1
Snellen et al. 2002, 2 de Vries et al. 1997, 3 Stanghellini et al.
1998, where $S \propto \nu^{-\alpha}$.}
         $$
\begin{tabular}{cccccccccccccc}

            \hline
            \noalign{\smallskip}
            1&2&3&4&5&6&7&8&9&10&11&12&13&14\\
$Name$ &other& $id$ & $m_{R}$ & $z$ & $pc/mas$ & $\theta$ & $L$ & $vlbi$ & $\alpha_{l}$ & $\alpha_{h}$ & $\nu_{m}$ & $S_{m}$ & $ref$ \\
& & & & && mas & pc &  & & & GHz & Jy & \\
            \noalign{\smallskip}
            \hline
            \noalign{\smallskip}

J0210+0419 &B0208+040& gl   & $>{24.1}$  & 1.5* &6.1    & 90  &          & cd &  &0.80  & 0.4 & 1.3 & 1 \\

J0323+0534 &4C+05.14& gl & 19.2    & 0.1785 &2.7   &180 &     490 &  cd&  &0.85   & 0.4 & 7.1 & 1 \\

J0433$-$0229 & 4C$-$02.17&gl & 19.1  & 0.530 &  5.1  & 80 & 408          & cd  &  &0.52    & 0.4 & 3.0 & 1 \\

J0913+1454 & B0910+151&gl  & $22.9$ & 0.47* & 4.9   & 80  &            & cd &  &0.75  & 0.6 & 1.1& 1 \\

J1057+0012& B1054+004&gl   & $22.3$ & 0.65* & 5.5  & 80? &            & cj &     &0.67  & 0.4   & 1.6  & 1\\

J1109+1043& B1107+109&gl   & 22.6   & 0.55*  & 5.2   & 60  &            & cd  &  & 0.94  & 0.5 & 2.4& 1 \\

J1135$-$0021 &4C$-$00.45& gl  & 21.9  & 0.975 & 6.0 & 120 &  720          & cd  &   &0.78   & 0.4 &2.9 &1  \\

J1203+0414& B1200+045& Q   & 18.8  & 1.221 &  6.1     & 75 &   458      &  cj &  &0.45   & 0.4 &1.4 &1  \\

J1352+0232 &B1349+027& gl  & 20.0  &  0.607  & 5.4   & 170  & 918           &  cd&  &0.58   & 0.4 & 2.0 & 1 \\

J1352+1107 & 4C+11.46&gl  & $21.0$ & 0.891&  5.9 & 50 &   295         & cd  &  &1.03   & 0.4 & 3.6 & 1 \\

J1600$-$0037 & B1557$-$004 &gl &        &      &        &  50 &            & cd/cj    &  &1.17   & 1.0 & 1.2 & 1 \\

J1648+0242 &4C+02.43& gl   & $22.1$  & 0.824&  5.8  &    &           &  n    &   &   &0.4  &3.4 & 1 \\

J2058+0540 & 4C+05.78&gl   & 23.4  & 1.381 & 6.1   &160  &     970    & cd &   &0.95   & 0.4 &3.1  & 1 \\

J2123$-$0112 & B2121$-$014 & gl   & 23.3  & 1.158 & 6.1 & 80     &   488     & cso & -0.56 & 0.75  & 0.4 & 2.0 & 1 \\

J2325$-$0344 &B2322$-$040 & gl  & 23.5 &  1.509  &  6.0     &75&450 & cso  & -0.42 & 0.75  & 1.4 & 1.2 & 1 \\

J0917+1113 & B0914+114 & EF   &      &       &             & 190 &            & cso  & -0.1  & 1.6   & 0.3 & 2.3 & 3 \\

J1753+2750 & B1751+278 & gl  & 21.7  & 0.86* & 5.9        & 50   &              & cj  & -0.27 & 0.57  & 1.4 &0.6 &2 \\

J1826+2708 & B1824+271 & gl  & 22.9  &       &             & 45  &            &  cso & -0.39 & 0.75  & 1.0 & 0.4 & 2 \\

J2325+7917 & B2323+790 & gl   & 19.5V &       &             & 32  &            &  cd  &-0.3  &0.75   &1.4  &1.2 & 2 \\

            \noalign{\smallskip}
            \hline
        \end{tabular}{}
         \label{tb1}
         $$
   \end{table*}

\section{VLBI structure and spectra of GPS sources}

The 1.6 GHz VLBI observation was carried out on 2006 March 3 using
the MK5 recording system with a bandwidth of 32 MHz and sample
rate of 256 Mbps. The EVN antennae in this experiment were
Effelsberg, Westerbork, Jodrell, Medicina, Noto, Onsala, Torun,
Hartebeesthoek, Urumqi, and Shanghai. Snapshot observations of 19
sources (Table~\ref{tb1}) in total of 24 hours were made. OQ208
was observed as a calibrator. The Astronomical Image Processing
System (AIPS) has been used for editing, a-priori calibration,
fringe fitting, self-calibration, and imaging of the data.

The results show that 12 out of 15 sources from Snellen (2002)
sample exhibit compact doubles, 2 sources exhibit core-jet
structure, and J1648+0242 is totally resolved out. For 4 others
sources in Table~\ref{tb1}, 3 show compact doubles, one shows a
core-jet. The last six sources in Table~\ref{tb1} were also
observed at 2.3/8.4 GHz and/or 5 GHz before this 1.6 GHz run. We
summarize the VLBI structure of 19 sources in Table~\ref{tb1}. Our
results show that about 80\% of the GPS galaxies exhibit a compact
double or CSO like structure. The quasar J1203+0414 tends to show
a core-jet. We have identified 4 CSOs from the sources which
observed at 4 frequencies, according to their symmetric and steep
spectrum of mini lobes. In the following we show the images and
spectra of two sources as examples.

\subsection{PKS 0914+114}

\begin{figure}
     \includegraphics[width=45mm,height=70mm,angle=270]{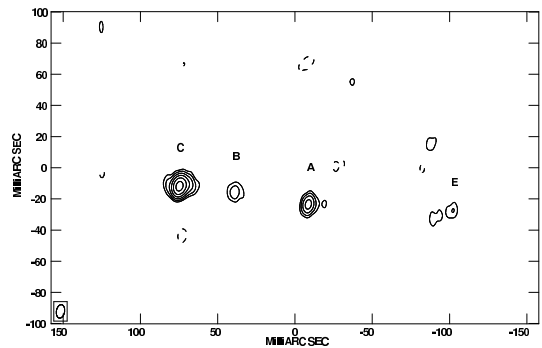}
     \caption{0914+114 at 2.3 GHz, the restoring beam is
        $10.4\times7.3$ mas with PA $-15.5^{\circ}$. The peak is 164 mJy/beam, the first contour
        is 2 mJy/beam. The contour levels here and below increase by a factor of 2.}
      \label{fig1}
   \end{figure}

\begin{figure}
     \includegraphics[width=45mm,height=70mm,angle=270]{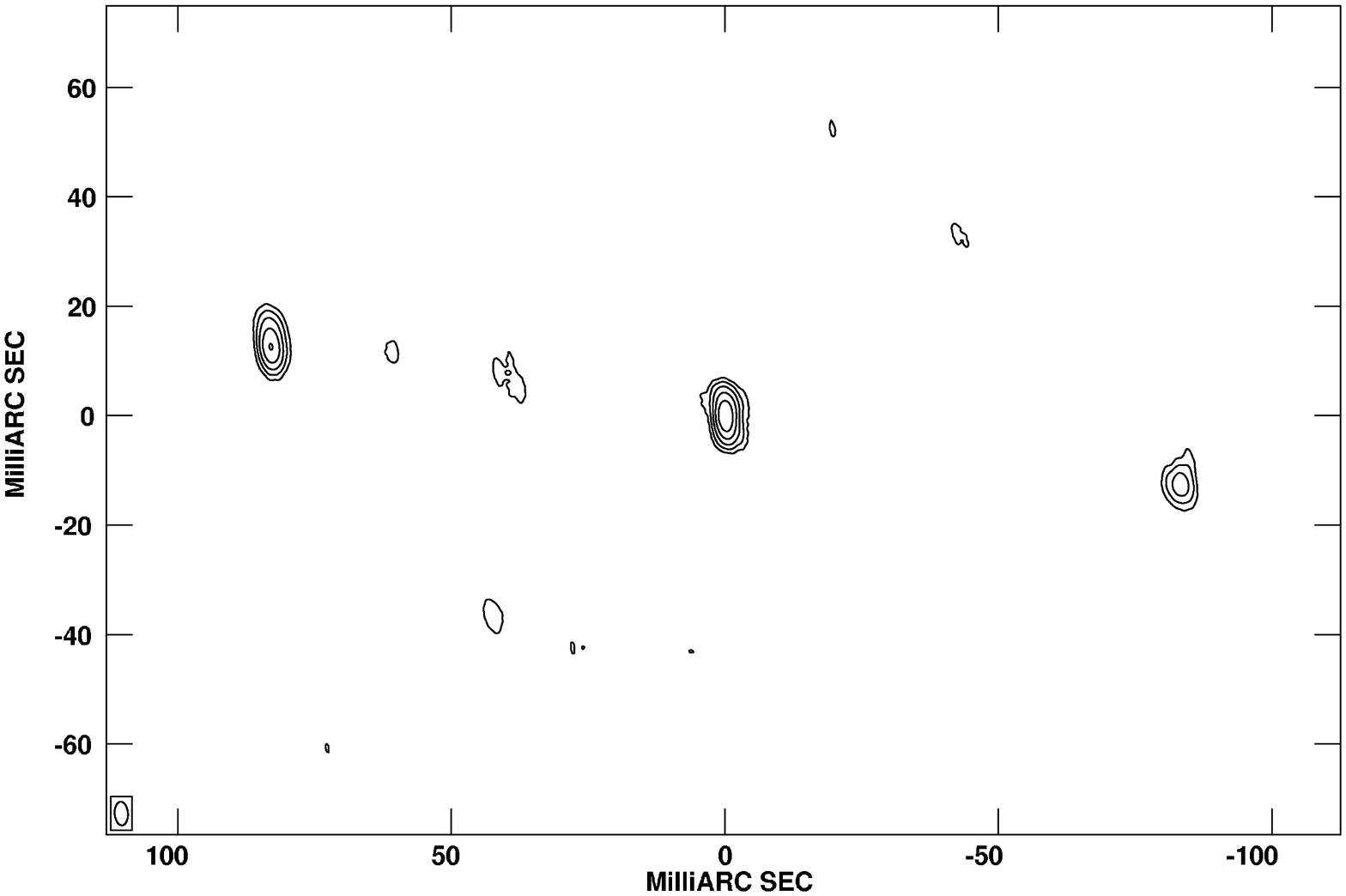}
     \caption{0914+114 at 5 GHz, the restoring beam is
        $4.3\times2.4$ mas with PA $3.9^{\circ}$. The peak is 30.7 mJy/beam, the first contour
        is 1 mJy/beam.}
      \label{fig2}
   \end{figure}

\begin{figure}
     \includegraphics[width=45mm,height=70mm,angle=270]{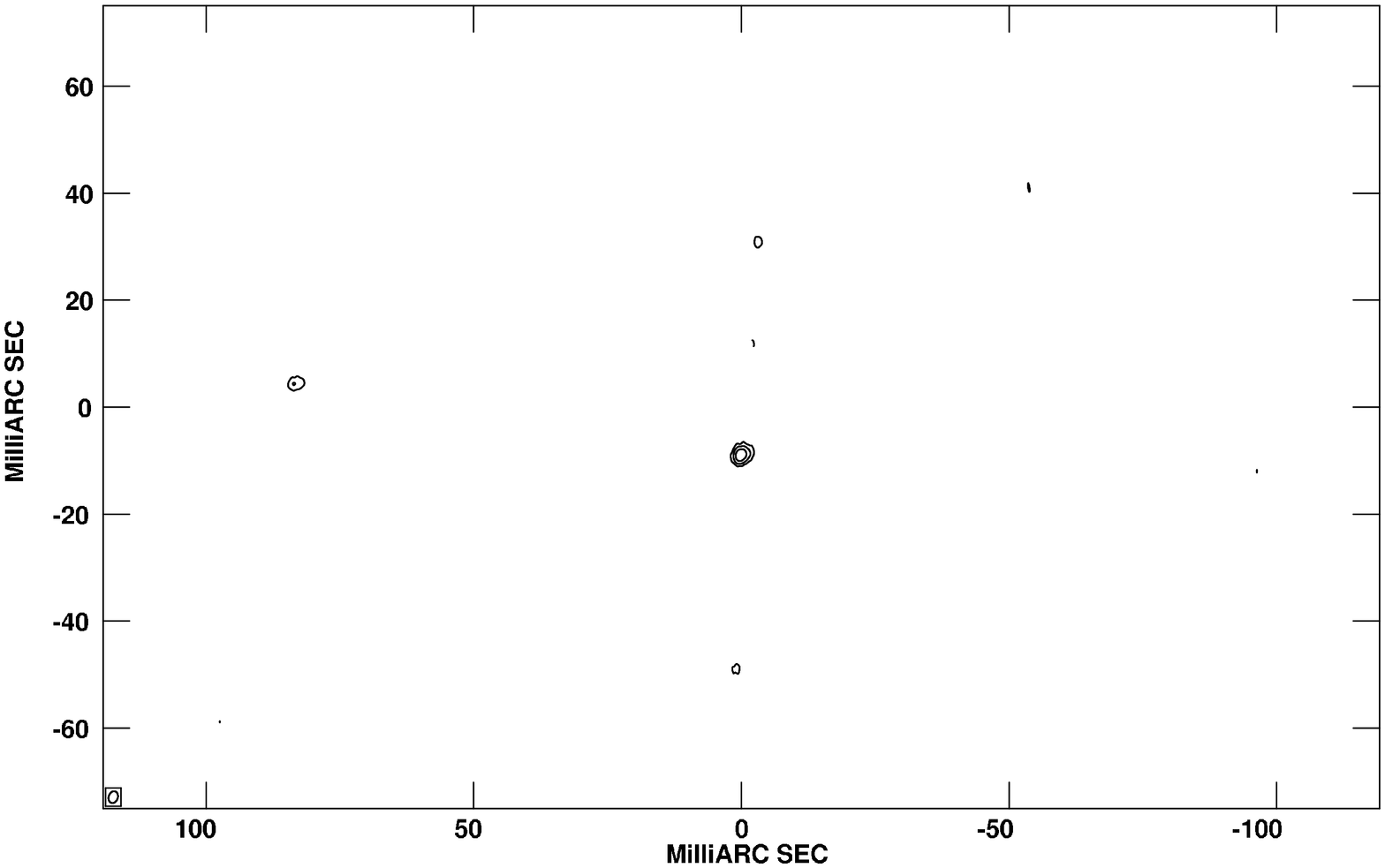}
     \caption{0914+114 at 8.4 GHz, the restoring beam is
        $2.2\times1.7$ mas with PA $-18.5^{\circ}$. The peak is 11.1 mJy/beam, the first contour
        is 1.5 mJy/beam.}
      \label{fig3}
   \end{figure}

The source has a GPS spectrum (Stanghellini et al. 1998), recently
it has been optically identified as an empty field (Labiano et al.
2007). In the 1.6 GHz VLBI image, it shows a core component A,
one-side jet B and two lobes C, E (Liu et al. 2007). Considering
the source structure at 1.6 GHz, we have re-processed the data at
2.3, 5, 8.4 GHz with careful calibration and `clean' in AIPS, the
component E has been restored at 2.3, 5 GHz, and component C is
restored at 8.4 GHz, as shown in Fig.~\ref{fig1}, Fig.~\ref{fig2},
Fig.~\ref{fig3}.

\begin{table}

\small \caption{Flux densities (mJy) of VLBI components, the value
with * is an upper limit. The error is $\sim$10\% of flux
density.}

\label{tb2}
\begin{tabular}{@{}crrrrr@{}}

\hline
Source &comp & $S_{1.6}$ & $S_{2.3}$ & $S_{5}$ & $S_{8.4}$\\

\hline

0914+114&A & 50 & 38 & 52  & 19\\
&B         & 65& 14 & 3 & 1*\\
&C         & 360 & 186 & 31 & 6 \\
&E         & 40& 17 & 10& 1*\\
2121$-$014&A & 594 &353 & 196 &113 \\
&B         &  10*& 120  & 14  & 4*\\
&C         & 415 & 237 & 107  & 29 \\

 \hline
\end{tabular}

\end{table}

 \begin{figure}
     \includegraphics[width=7cm]{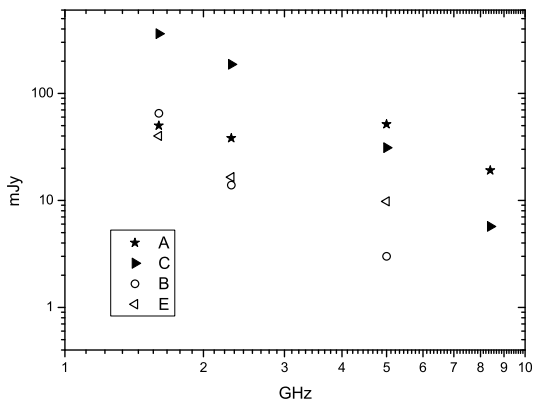}
     \caption{Component spectra of 0914+114 at 1.6, 2.3, 5, and 8.4 GHz. Total flux density peaks at 0.3 GHz.}
      \label{fig4}
   \end{figure}

The component spectra (Fig.~\ref{fig4}) from the VLBI images at
1.6, 2.3, 5, and 8.4 GHz, show that the core A has a flatter
spectrum than other components. The data are listed in
Table~\ref{tb2}. Based on its structure and spectra, we classify
this GPS source as a CSO.

\subsection{PKS 2121$-$014}

\begin{figure}
     \includegraphics[width=7cm]{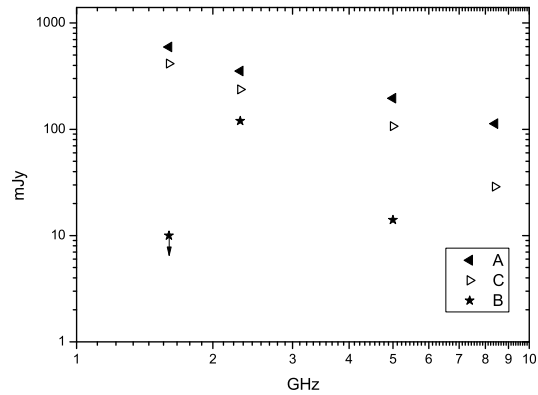}
     \caption{Component spectra of 2121$-$014 at 1.6, 2.3, 5, and 8.4 GHz. Total flux density peaks at 0.4 GHz.}
      \label{fig5}
\end{figure}

   \begin{figure}
     \includegraphics[width=45mm,height=70mm,angle=270]{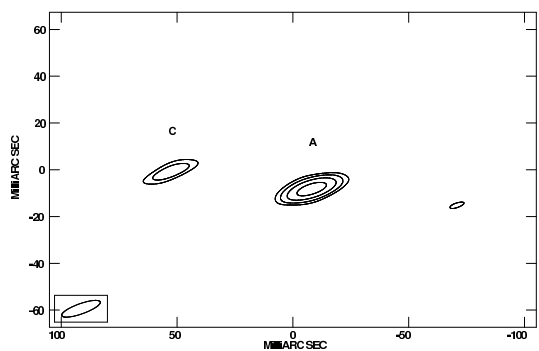}
     \caption{2121$-$014 at 8.4 GHz, the restoring beam is
        $17.6\times4.4$ mas with PA $-70.3^{\circ}$. The peak is 82.7 mJy/beam, the first contour
        is 7 mJy/beam.}
      \label{fig6}
   \end{figure}

This GPS source is hosted by a galaxy at a redshift of 1.158.
Using the VLBI results at 1.6, 2.3, 5, and 8.4 GHz we obtained the
spectra for each component (Fig.~\ref{fig5}), the two lobes A, C
show steep spectra, in which we have reprocessed the 8.4 GHz data
as shown in Fig.~\ref{fig6} with careful calibration and `clean'
in AIPS. As discussed in Xiang et al. (2006), the two components
A, C in the 8.4 GHz image (Xiang et al. 2005) might be swapped due
to worse phase calibration, so the new image Fig.~\ref{fig6} is a
correction. The component B as seen at 2.3, 5 GHz (Xiang et al.
2005, 2006) was not detected in this 1.6 GHz observation, assuming
an upper limit of 10 mJy set by $3\sigma$ in the image, it has an
inverted spectrum which is a sign of a jet or probably an absorbed
core. Based on its double structure and steep spectra of lobes, we
classify this GPS source as a CSO.

\section{Flux variability of GPS sources}

It is reported that a lot of GPS sources show long term
variability (Torniainen et al. 2005) especially at high radio
frequency. In order to check this also at centimeter wavelength,
in 2007 July, we carried out flux density measurements at 4.85 GHz
for a large sample of 172 GPS sources edited by Labiano et al.
(2007). By comparing these data with the 87GB and PMN data at the
same frequency, we obtained the flux variations of the GPS sources
between the two epoches. Because low source declination and/or bad
Gaussian fits, only flux densities of 121 sources were obtained.

Flux densities were determined with `cross - scans' in azimuth and
elevation, fourfold in each coordinate. This enables us to check
the pointing offsets in both coordinates. A Gaussian fit was
performed on each sub-scan, and after applying a correction for
pointing offsets the amplitudes of both AZ and EL were averaged.
Then we correct the measurements for the antenna gain, and finally
the flux densities were scaled to 7.55 Jy of 3C286.

\begin{figure}
     \includegraphics[width=7cm]{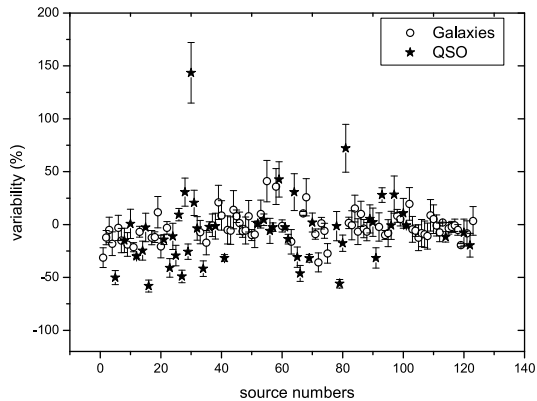}
     \caption{Flux variation (\%) at 4.85 GHz, by comparing the data measured in 2007
at Urumqi telescope with 87GB/PMN data for 121 GPS sources from
Labiano et al. sample.}
      \label{fig7}
   \end{figure}

As in Fig.~\ref{fig7}, only 9 out of 73 GPS galaxies (12\%) show
flux variation higher than 10\% ($>2\sigma$); at the same time 21
out of 48 quasars (44\%) show variability higher than 10\%
($>2\sigma$). The result indicates that GPS quasars are much more
variable than GPS galaxies. The variable sources are listed in
Table~\ref{tb3}, columns 1 to 6 are source name (1950), optical
identification, 4.85 GHz flux density measured at Urumqi, 4.85 GHz
flux density from 87GB or PMN data, and flux variation and the
references (1: Gregory \& Condon 1991; 2: Griffith et al. 1994; 3:
Becker et al. 1991). Sources in Table~\ref{tb1} are also in
Labiano sample, of them two sources show variability, the core-jet
source 1054+004 shows a variation of (11.4$\pm$3.6)\%, and
1557$-$004 shows a variation of (13.4$\pm$6.9)\% just at $2\sigma$
level.

\begin{table}

\small \caption{Flux density variation ($>2\sigma$) of the GPS
sources.}

\label{tb3}

\begin{tabular}{@{}crrrrr@{}}

\hline

Source & id & $S_{ur}$ & 87GB/PMN & variation & ref\\


\hline

0000+212 & gl &242$\pm$6 &352$\pm$47  & -31.3$\pm$9.3 &1\\

0039+230 &Q & 820$\pm$6 &1645$\pm$220 &-50.2$\pm$6.7 &1\\

0207$-$224 & gl &486$\pm$1  &618$\pm$34 &-21.4$\pm$4.3 &2\\

0237$-$233 & Q & 2541$\pm$5 &3630$\pm$99& -30$\pm$1.9 &2\\

0248+430 & Q &1065$\pm$6 &1414$\pm$169 &-24.7$\pm$9 &1\\

0354+231 & Q &137$\pm$2  &327$\pm$44  &-58.1$\pm$5.7 &1\\

0405$-$280& gl &487$\pm$4  &553$\pm$31  & -11.9$\pm$5.0 &2\\

0434$-$188 & Q &934$\pm$4 &1089$\pm$58  & -14.2$\pm$4.6  &2\\

0457+024 & Q & 996$\pm$5  &1689$\pm$253  &  -41.0$\pm$8.8 &1 \\

0507+179 & Q & 549$\pm$5 & 777$\pm$106 & -29.3$\pm$9.7 &1\\

0621+446& Q & 188$\pm$3 & 369$\pm$43 & -49.1$\pm$6.0 &1\\

0636+680& Q & 371$\pm$17 & 499$\pm$43 & -25.7$\pm$7.3 &1\\

0642+449& Q & 2900$\pm$7 & 1191$\pm$140 & 143.5$\pm$28.6 &1 \\

0711+356& Q & 525$\pm$2 & 901$\pm$113 & -41.7$\pm$7.3 &1\\

0858$-$279& Q & 1516$\pm$12 & 2216$\pm$99 & -31.6$\pm$3.1 &2\\

1054+004 & gl & 351$\pm$6 & 396$\pm$23 & -11.4$\pm$3.6 &2\\

1200+045& gl & 721$\pm$5 & 511$\pm$71 & 41.1$\pm$19.6 &1\\

1315+415& gl & 242$\pm$3 & 178$\pm$22 & 36.0$\pm$16.9 &1\\

1333+459& Q & 853$\pm$7&  598$\pm$70& 42.6$\pm$16.7 &1\\

1354$-$174& Q & 870$\pm$5 & 1009$\pm$54 & -13.8$\pm$4.6 &2\\

1427+109 & Q& 855$\pm$5 & 1236$\pm$171 &  -30.8$\pm$9.6 &1\\

1502+036& Q & 533$\pm$3 & 991$\pm$138 & -46.2$\pm$7.5 &1\\

1519$-$273& Q & 1250$\pm$18 & 1835$\pm$96 & -31.9$\pm$3.7 &2\\

1543+005& gl & 839$\pm$5 & 1306$\pm$182 & -35.7$\pm$9.0 &1\\

1600+335& gl & 1492$\pm$7 & 2051$\pm$262 & -27.3$\pm$9.3 &1\\

1622+665& gl & 229$\pm$6 & 520$\pm$46 & -55.9$\pm$4.1 &1\\

1645+635& Q & 366$\pm$6 & 444$\pm$41 & -17.6$\pm$7.7 &1\\

1648+015& Q & 787$\pm$5 & 457$\pm$60 & 72.2$\pm$22.6 &3\\

2019+050& Q & 468$\pm$6 & 684$\pm$95 & -31.6$\pm$9.5 &1\\

2126$-$158& Q & 1517$\pm$12 & 1186$\pm$63 & 27.9$\pm$6.9 &2\\

 \hline

\end{tabular}

\end{table}

\section{Summary and discussion}

Our VLBI results show that 80\% of the GPS galaxies exhibit a
compact double or CSO-like structure. However, for 13 sources
which were only observed at 1.6 GHz further VLBI observations are
needed for classifying their structure type in detail. Combining
the 1.6, 2.3/8.4, and 5 GHz VLBI observations for other 6 sources,
4 CSOs are classified according to their steep spectrum of double
lobes. The GPS galaxies are dominated by jet/lobe emission, with a
hidden or weak core (it was not detected because of sensitivity in
most of our sources), suggesting they are at large angles to the
line of sight. While GPS quasars may be at moderate angles to the
line of sight (Stanghellini et al. 2001).

Only 12\% of GPS galaxies show flux variation higher than 10\%,
while 44\% of GPS quasars show variability higher than 10\% in
passed 20 years. The result indicates that most GPS galaxies are
stable in flux density, and the GPS quasars are much more variable
than GPS galaxies. The variable GPS quasars often show compact
core-jet, e.g. the quasar 0642+449 which has 140\% flux variation
in Fig.~\ref{fig7} is an extremely compact core-dominated one at a
redshift of 3.4. Torniainen et al. (2005) found that 54\% of GPS
sources show flux variability in long term (with fractional
variability index $>3$). In our result, 25\% of GPS sources show
$>10\%$ variation, the difference may be due to different samples
and the definition of variability, we simply compared the flux
densities in two epoches.

GPS galaxies which show comparable double lobes can be interpreted
as type-2 AGN in the framework of unified scheme. GPS quasars,
which show a core-jet structure and variability, could be mostly
type-1 AGN.

\acknowledgements{} This work was supported by the National
Natural Science Foundation of China (NSFC) under grant No.10773019
and the 973 Program of China under grant No.2009CB824800.


\end{document}